\documentstyle[12pt,aasms4,flushrt]{article}

\begin{document}

\title{Optical Emission Lines \\from Warm Interstellar Clouds -
a Decisive Test of the \\ Decaying Neutrino Theory.}

\author{D.W. Sciama}

\affil{SISSA and ICTP, Trieste \\
Via Beirut 2-4, 34013 Trieste, Italy\\ 
Nuclear and Astrophysics Laboratory, University of Oxford, Keble 
Road, Oxford \ OX1 3RH, England}

\begin{abstract}
Recently developed instruments such as the Taurus Tunable Filter and WHAM 
should be able to detect some or all of the optical emission lines
H$\alpha$ , [OI] $\lambda$6300, [SII] $\lambda$6717, [NI] $\lambda$
5200 and [NII] $\lambda$6584 from warm interstellar clouds such as those
observed by Spitzer $\&$ Fitzpatrick\markcite{1} (1993) (SF) along the
line of
sight to the halo star HD93521. The strengths of these lines should
resolve the debate as to whether the free electrons, which SF held
responsible for the
observed excitation of CII in the clouds, are located mainly in the skins
of the clouds or in their interiors.
If the free electrons are indeed mainly located in the cloud interiors,
then the substantial electron density derived by SF, and its constancy
from cloud to cloud for the slow-moving clouds, when combined with their
opacity to Lyman continuum radiation, lend strong support to the
decaying neutrino theory for the ionisation of the interstellar medium
(Sciama 1990, 1993 a, b, 1997)\markcite{2}\markcite{3}\markcite{4}\markcite{5}.
If the [OI] and [NI] lines are relatively strong but the [NII] line is
weak, then this would lend further, decisive, support to this theory,
since decay photons are unable to ionise N, although its ionisation
potential is only 0.9 eV greater than that of H. 
\end{abstract}


\section{Introduction}

The recent development of sensitive instruments such as the Taurus Tunable
Filter (Bland-Hawthorn $\&$ Jones 1998)\markcite{6} and WHAM (Reynolds et
al 1998a)\markcite{7} for
observing faint optical emission lines may make possible the detection of
such lines from individual warm interstellar clouds. It would be of
particular interest to study in this way the clouds observed by Spitzer
$\&$ Fitzpatrick (1993)\markcite{1} (SF) along the line of sight to the
halo star HD
93521. Their HST observations of the spectrum of this star revealed the
presence of nine separate warm ($T \simeq$ 6000 K) absorbing clouds. Four
of
these clouds were slowly moving relative to the sun, the modulus of their
heliocentric velocities being less than 19 km sec$^{-1}$. The remaining
five clouds had higher relative velocities, up to -66.3 km sec$^{-1}$. One
of the absorbing ions which SF observed in each cloud was CII*
($J = 3/2$), whose excitation they attributed to collisions
betwen CII ions and free electrons. They were able to derive from their
data on the column density N(CII*) the mean electron density $n_e$ in
each cloud and found that $n_e$ is (i) substantial ($\sim 0.05 - 0.1$
cm$^{-3}$ in the slow clouds depending on the assumed abundance of C) and
(ii) the same in each of the slow clouds to a precision $\sim 10$ per
cent.
They argued that these free electrons are mainly located in the interiors
of the clouds rather than in their skins (as later proposed by
Domg\"orgen and Mathis 1994)\markcite{8}.

Further arguments in favour of the interior location of the free electrons
were given by Sciama (1993 b, 1997)\markcite{4}\markcite{5} who also
claimed that, because each of
the clouds has a column density of H I exceeding $5.6 \times 10^{18}$
cm$^{-2}$, and so is opaque to hydrogen-ionising photons near the Lyman
edge, properties (i) and (ii) strongly suggest that the free electrons in
the clouds are mainly produced by ionising photons emitted by "dark"
matter neutrinos in the clouds (Sciama 1990, 1993
a)\markcite{2}\markcite{3}.
The aim of this letter is to point out that, if the strengths of the
emission lines H$\alpha$, [OI] $\lambda$6300, [SII] $\lambda$6717, [NI]$
\lambda$5200 and [NII] $\lambda$6584 could be measured in the clouds, it should
be
possible to settle these questions conclusively.   

\section{Optical Emission Lines from Warm Interstellar Clouds}

With the exception of H$\alpha$, all the emission lines listed above
require the emitting atoms or ions to be excited by electron collisions.
Accordingly, for the lines to be observable the clouds concerned would
have to be warm  ($T > 3000$ K). In fact the SF clouds have
$T \sim 6000$ K, as
judged by the widths of their 21 cm emission lines and their various
absorption lines. The strengths of the optical emission lines, and also of
the absorption line of CII*, are each determined (except for H$\alpha$)
by
the product of the column density of the emitting atom or ion, the mean
electron density, a function of the temperature through each cloud, and
the collision cross-section. As is well known from many observational and
theoretical studies, it so happens that, for the elements O, S, and N,
their interstellar abundances relative to H, when combined with a
temperature $\sim 10^4$ K and their excitation cross-section
$\sim 10^{-15}$ cm$^2$ (Osterbrock 1989)\markcite{9}, lead to emission
fluxes in the
lines comparable to, but somewhat less than, the flux of H$\alpha$ from
the same diffuse region of the ISM, if the dominant ionisation state of
the emitting atom or ion is the one which produces the lines concerned.
This
simple coincidence underlies our proposed use of these emission lines to 
test the decaying neutrino theory. We now consider each of these emission
lines, and the absorption line of CII*, in turn.

\noindent
(i) CII*

We assume for simplicity that each cloud has a uniform total hydrogen
density $n_H$ and temperature T, and that the hydrogen in the skin of each
cloud is completely ionised. Then we have from SF that

\begin{equation}
{\rm N} ({\rm CII}^{*}) = \frac{\xi_C n_H \chi_c}{T^{1/2}}
(n_H
l_s + n_e l_c), 
\end{equation}

\noindent where $\xi_C$ is the abundance of C relative to H, $\chi_c$ is
the
collision strength, $l_s$ and $l_c$  are the total thicknesses of the
skins
and
the interior of a cloud, and $n_e$ is the electron density in the
interior of a cloud. We have assumed that CII is the dominant ionisation
stage of C throughout each cloud.
SF observed that, for the four slowly-moving, and presumably shock-free,
clouds N(CII*) $\propto$ N(HI). The natural conclusions from this
observed proportionality are that (a) $n_H l_s \ll n_e l_c$, (b)
inside each of these clouds $n_{HI} \gg n_e$,  and (c)
$n_e$ is the same in each of these clouds. In that case

\begin{equation}
{\rm N} ({\rm CII}^{*}) = \frac{\xi_C \chi_c}{T^{1/2}} n_e
{\rm N}({\rm HI}).
\end{equation}

Since $\chi_C$ and T are known one can derive $n_e$ from this relation if
$\xi_C$ can be determined. Unfortunately the CII absorption line is
saturated in each cloud, and SF had to make an assumption about the value
of $\xi_C$. As we shall see below, this uncertainty can be avoided if the
strength of the [SII] emission line can be measured, since N(SII) was
directly observed in each cloud.

\noindent
(ii) H$\alpha$

We have, for case B,

\begin{equation}
{\rm H}\alpha = {\case{1}{2}} \alpha (n_H^2 l_s + n^2_e l_c),
\end{equation}

\noindent where $\alpha$ is the appropriate recombination coefficient. If
the free electrons were mainly in the skin of each cloud we would then have

\begin{eqnarray}
{\rm H}\alpha & = & {\case {1}{2}} \alpha n_H^2 l_s \\
& = & {\case{1}{2}} \frac{\alpha T^{1/2}}{\xi_C \chi_c} {\rm N}({\rm
CII}^*).
\end{eqnarray}

\noindent If $\xi_C$ is solar, $T = 6000$ K, and N(CII*) $= 2.3
\times 10^{13}$
cm$^{-2}$
(as it
is in cloud number 6) we would obtain H$\alpha$ = 2 x 10$^5$ photons
cm$^{-2}$ sec$^{-1}$ for this cloud alone.

On the other hand, if the free electrons were mainly in the interior of
each cloud we would have

\begin{equation}
{\rm H}\alpha = {\case{1}{2}} \frac{\alpha T^{1/2}}{\xi_C
\chi_c} \frac{n_e}{n_H}{\rm N}({\rm CII}^*),
\end{equation}

\noindent which is less than the previous value by the (small) factor
$n_e$/$n_H$.
It follows that a measurement of H$\alpha$ from the clouds would
determine
the location of the free electrons and would constrain $n_e$/$n_H$ if the
electrons are found to be mainly in the interior. There exist
unpublished measurements and 
upper limits on H$\alpha$ in this direction obtained by
Reynolds (private communication), and these observations were used by Sciama
(1997)\markcite{5} to deduce that the free electrons are indeed mainly
in the
interiors of the clouds, and that $n_e/n_H \sim 1/3.5$   or  
$1/7$, depending on whether the abundance of C is
assumed to be
solar or half-solar. It would be desirable to confirm this argument by
obtaining definite values of H$\alpha$ in each cloud from the new
instruments.  It might also be possible to observe limb brightening
if the angular resolution permits (Bland-Hawthorn, private communication).

\noindent
(iii) [O I] $\lambda$6300

This line has recently been detected in the diffuse ISM by WHAM (Reynolds
et al
1998 b)\markcite{10}. Its ratio to H$\alpha$ is so low ($\sim$0.01-0.04)
that Reynolds
et al concluded that in the regions concerned H must be highly ionised,
since the ionisation level of O is tied to that of H by charge exchange
(Field $\&$ Steigman 1971)\markcite{11}. Since the line is only
detectable from regions
which contain appreciable densities of both neutral O and free electrons,
it is ideal for testing the claim that the SF clouds are mainly neutral in
their interiors but also contain an appreciable density of free
electrons. In that case we would have (Reynolds 1989)\markcite{12}

\begin{equation}
[{\rm OI}] = 2 \times 10^{-9} T_4^{0.45}  \exp ( -2.28 / T_4 ) 
\xi_O n_e
{\rm N}({\rm HI}).
\end{equation}

\noindent For T$_4$ = 0.6, $\xi_O = 8.5 \times 10^{-4}$, n$_e$ = 0.05 
cm$^{-3}$ and N(HI)
$= 2 \times 10^{19}$ cm$^{-2}$ as found for cloud 6, one would obtain

\begin{equation}
[{\rm OI}] = 2.8 \times 10^4  {\rm cm}^{-2} {\rm sec}^{-1}
\end{equation}

\noindent for this cloud alone. A flux of this order would be measurable by the
new
instruments if the geocoronal [OI] line can be avoided. This might
require confining the observations to the fast clouds. If a flux of this
order is found it would show conclusively that $n_e$ is substantial in the
interiors of the clouds. The observed flux would also determine
$\xi_O / \xi_C$ for each cloud.

\noindent
(iv) [{\rm SII}] $\lambda$6717

Since S is expected to be mainly in the form SII in both the skin and the
interior of each cloud, one would have (Reynolds 1989)\markcite{12}

\begin{equation}
[{\rm SII}] = 7.3 \times 10^{-8} T_4^{-0.5} \exp (-2.14 / T_4) n_e {\rm 
N(SII)},
\end{equation}

\noindent if one assumes that, because SF found that N(SII) $\propto$ 
N(HI),
most of
the
SII is in the interior of each cloud. For $T_4 = 0.6$,  $n_e = 0.05$
cm$^{-3}$ and N(SII) $= 3.2 \times 10^{14}$ cm$^{-2}$ (as it is in 
cloud
6), one would have

\begin{equation}
[{\rm SII}] = 4.1 \times 10^4 {\rm cm}^{-2} {\rm sec}^{-1}
\end{equation}

\noindent for this cloud alone. Again this is a measurable flux with the new
instruments.
Since N(SII) was measured by SF in each cloud, the flux of [SII] can be
used to derive the value of n$_e$ in each cloud, without having to make
an assumption about the abundance of SII which, in fact, SF found to be
solar from the observed ratio ${\rm N(SII)}/{\rm N(HI)}$. 
This value of n$_e$ could then be used to evaluate both
the ionisation rate of H and the abundance of C and O in each cloud.

\noindent
(v) [NI] $\lambda$5200

Although the ionisation level of N is no longer thought to be tied to that
of H by charge exchange (Osterbrock 1989)\markcite{9}, one would still
expect that in
the interiors of the clouds N would be mainly in the form NI. If both N
and O have solar abundances in the clouds one has (Silk 1970)\markcite{13}

\begin{equation}
\log \frac{\rm [OI]}{\rm [NI]} = 0.29 + \frac{0.16}{T_4}.
\end{equation}
Hence for cloud 6 one would have

\begin{equation}
[{\rm NI}] = 7.8 \times 10^3 {\rm cm}^{-2} {\rm sec}^{-1}.
\end{equation}

If this line could be detected one would again be able to test, as with
the [OI] line, whether the free electrons are mainly located in the
interiors of the clouds. In addition one would be able to determine the
abundance of N in the clouds.

\noindent
(vi) [NII] $\lambda$6548

A key question for the decaying neutrino theory is whether the decay
photons can ionise N (ionisation potential 14.5 eV) as well as H (13.6
eV). As pointed out by Sciama (1995)\markcite{14}, if the extragalactic
hydrogen-ionising flux at zero red shift is less than $\sim 10^5$
photons cm$^{-2}$sec$^{-1}$ (as claimed by Vogel et al 
(1995)\markcite{15} and by
Donahue et al (1995)\markcite{16} from their H$\alpha$ observations of
intergalactic
neutral H clouds) then the energy of a decay photon must be less than
$\sim 13.7$ eV, so that N would not be ionised by decay photons. If,
however, N is ionised in clouds by the same mechanism as is H, then a
comparison of the photoionisation and recombination rates of N and H
suggests that NII/NI would have the same order of magnitude as HII/HI. By
contrast  if the H in the interiors of the clouds is mainly ionised by
decay photons, than the ionisation level of N in the interiors of the
clouds would be much less than that of H.

One could determine NI/NII in the cloud interiors if one could measure
both
[NII] $\lambda$6584 and [NI] $\lambda$5200 in the clouds. This follows
because one has (Reynolds et al 1977)\markcite{17}

\begin{equation}
[{\rm NII}] = \frac{\exp(0.57/T_4)}{0.89 T_4}\frac{\rm NII}{\rm NI} [{\rm NI}].
\end{equation}
Hence for T${_4}$ = 0.6,

\begin{equation}
[{\rm NII}] = 4.85 \frac{\rm NII}{\rm NI}[{\rm NI}].
\end{equation}
If one finds that NII/NI has the same order of magnitude as
$n_e/n_H$ as determined by the H$\alpha$ data, then a
conventional mechanism for the ionisation of H would be indicated.
However, if NII/NI $\ll n_e/n_H$, then the decaying neutrino
theory would be confirmed.

\section{Conclusions}

We conclude that observations of optical emission lines from the warm
interstellar clouds  along the line of sight to the halo star HD93521
could 

(a) decide whether their free electrons are mainly located in the skins or
in the interiors of the clouds,

(b) provide a measurement of the mean electron density in each cloud free
of uncertain assumptions about element abundances,

(c) determine the abundances of the elements concerned in the clouds,

(d) determine whether NII/NI in the clouds is comparable to or
much less than HII/HI.

If the latter is found to be the case, this would provide decisive
evidence in favour of the decaying neutrino theory, whose photons are
unable to ionise N although its ionisation potential exceeds that of H by
only $0.9$ eV.

\acknowledgments{I am grateful to Ron Reynolds for his crucial help, and
to him and Joss Bland-Hawthorn for their comments on the manuscript.
I am grateful to the MURST for their financial support for this work.}


\begin{references}
\reference{6}Bland-Hawthorn, J., \& Jones, D. H. 1998, PASA, 15, 44
\reference{8}Domg\"orgen, H., \& Mathis, J. S. 1994, ApJ, 428, 647
\reference{16}Donahue, M., Aldering, G., \& Stocke, J. T. 1995, ApJ, 450, L45
\reference{11}Field, G. B., \& Steigman, G. 1971, ApJ, 166, 59
\reference{9}Osterbrock, D. E. 1989, Astrophysics of Gaseous Nebulae and 
Active Galactic Nuclei (Mill Valley: University Science Books)
\reference{12}Reynolds, R. J. 1989, ApJ, 345, 811
\reference{17}Reynolds, R. J., Roesler, F. L., \& Scherb, F. 1977, ApJ, 
211, 115
\reference{7}Reynolds, R. J., Tufte, S. L., Haffner, L. M., Jaehnig, K., 
\& Percival, J. W. 1998a, PASA, 15, 14 
\reference{10}Reynolds, R. J., Hausen, N. R., Tufte, S. L., Haffner, L. 
M. 1998b, ApJ, 494, L99
\reference{2}Sciama, D. W. 1990, ApJ, 364, 549
\reference{3}Sciama, D. W. 1993a, Modern Cosmology and the Dark Matter 
Problem (Cambridge: Cambridge Univ. Press)
\reference{4}Sciama, D. W. 1993b, ApJ, 409, L25
\reference{14}Sciama, D. W. 1995, ApJ, 448, 667
\reference{5}Sciama, D. W. 1997, ApJ, 488, 234
\reference{13}Silk, J. 1970, ApJ, 161, L37
\reference{1}Spitzer, L. \& Fitzpatrick, E. L. 1993, ApJ, 409, 299
\reference{15}Vogel, S. M., Weymann, R., Rauch, M., \& Hamilton, T. 1995,
ApJ, 441, 162
\end{references}
\end{document}